\documentclass[runningheads]{llncs}
\usepackage{graphicx}
\usepackage{caption}
\usepackage{subcaption}
\usepackage{multirow}
\usepackage[table,xcdraw]{xcolor}

\begin{document}

\title{Real vs Simulated Foveated Rendering to Reduce Visual Discomfort in Virtual Reality}
\titlerunning{Real vs simulated foveated rendering}
%

\author{Ariel Caputo\inst{1}\orcidID{0000-0002-6478-4663} \and Andrea Giachetti\inst{1} \orcidID{0000-0002-7523-6806} \and Salwa Abkal\inst{2} \and Chiara Marchesini\inst{2}  \and Massimo Zancanaro\inst{2,3} \orcidID{0000-0002-1554-5703}}
\authorrunning{Caputo et al.}
%
\institute{University of Verona, Department of Computer Science  \and University of Trento, Department of Psychology and Cognitive Science \and Fondazione Bruno Kessler }

\maketitle              

\begin{center}\fbox{\parbox[c]{0.5\textwidth}{
    {\centering\scriptsize preprint version - please cite as\\
    Caputo A., Giachetti A., Abkal S., Marchesini C., Zancanaro M.  (2021) Real vs Simulated Foveated Rendering to Reduce Visual Discomfort in Virtual Reality. In Proceedings of the 18th International Conference promoted by the IFIP Technical Committee 13 on Human–Computer Interaction - INTERACT 2021. August 30th - September 3rd, 2021, Bari, Italy}
    }}
\end{center}
\begin{abstract}
In this paper, a study aimed at investigating the effects of real (using eye tracking to determine the fixation) and simulated foveated blurring in immersive Virtual Reality is presented. Techniques to reduce the optical flow perceived at the visual field margins are often employed in immersive Virtual Reality environments to alleviate discomfort experienced when the visual motion perception does not correspond to the body's acceleration. Although still preliminary, our results suggest that for participants with higher self-declared sensitivity to sickness, there might be an improvement for nausea when using blurring. The (perceived) difficulty of the task seems to improve when the real foveated method is used.

\end{abstract}
\section{Introduction}
The perceptual mismatch in sensory cues between the vestibular and the visual systems is known to induce in human subjects the so-called motion sickness that may strongly affect users' experience in Virtual Reality (VR) where they navigate synthetic environments while being still \cite{kennedy2010research}. The visual discomfort can harm the diffusion and acceptance of immersive VR applications \cite{sagnier2020user}.

As discomfort is often related to the optical flow perceived at the margins of the field of view, a potential solution to reduce this effect is to reduce the amount of this optical flow. 
A reasonable way to do this consists of blurring or obscuring the images at the margins of the field of view (marginal blurring/FOV reduction). An approximation of foveated rendering can be obtained by assuming that the user's gaze is consistently directed along the virtual camera axis. Therefore, image quality is varied with the distance from the center of the viewpoint \cite{patney2016towards}.

Nowadays, low-cost eye-tracking devices are integrated into several Head Mounted Displays for Virtual Reality, making it possible to implement a truly foveated rendering, modifying the image's appearance far from the actual user's gaze point. A few recent papers \cite{pai2016gazesim,adhanom2020effect} proposed the use of eye-tracking to implement foveated rendering for the reduction of motion sickness. 

There is some empirical evidence on the positive effect of optical flow reduction techniques  (for example, \cite{adhanom2020effect},\cite{carnegie2015reducing}). Yet, the advantage of using a foveated approach by tracking the user's gaze rather than a simulated one (assuming that the gaze point coincides with the image center) has not been thoroughly investigated, at least for the blurring technique. 

\begin{figure}[t]
\centering
\begin{subfigure}{0.48\textwidth}
\includegraphics[width=\textwidth]{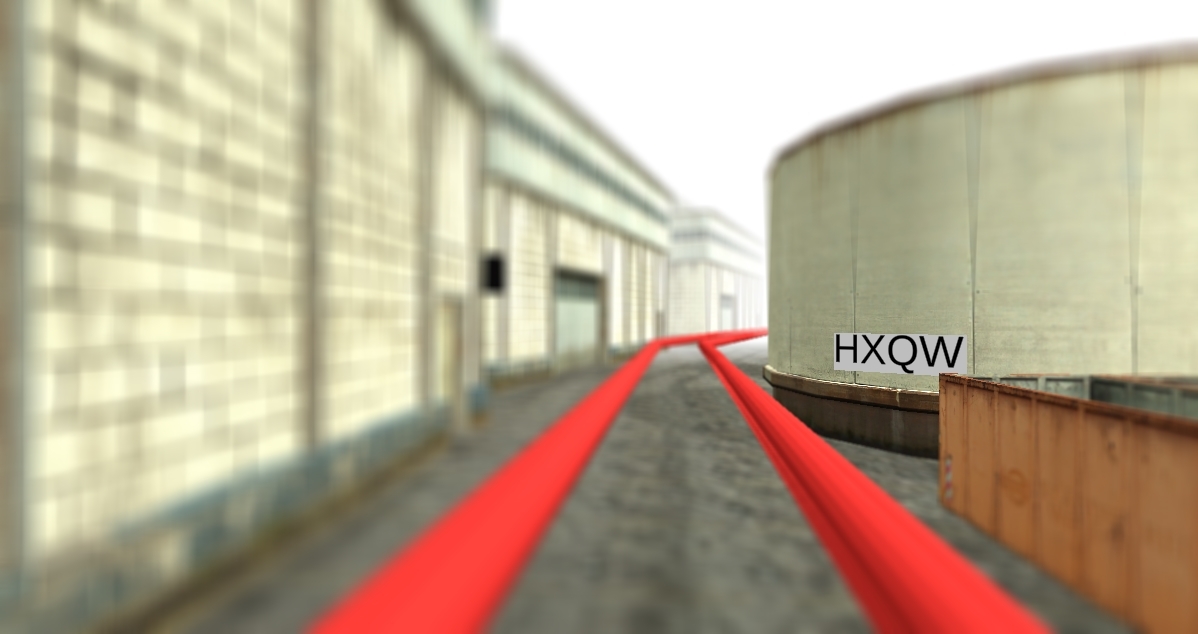}
\caption{} 
\label{fig:fovblurrec}
\end{subfigure}
\begin{subfigure}{0.48\textwidth}
\includegraphics[width=\textwidth]{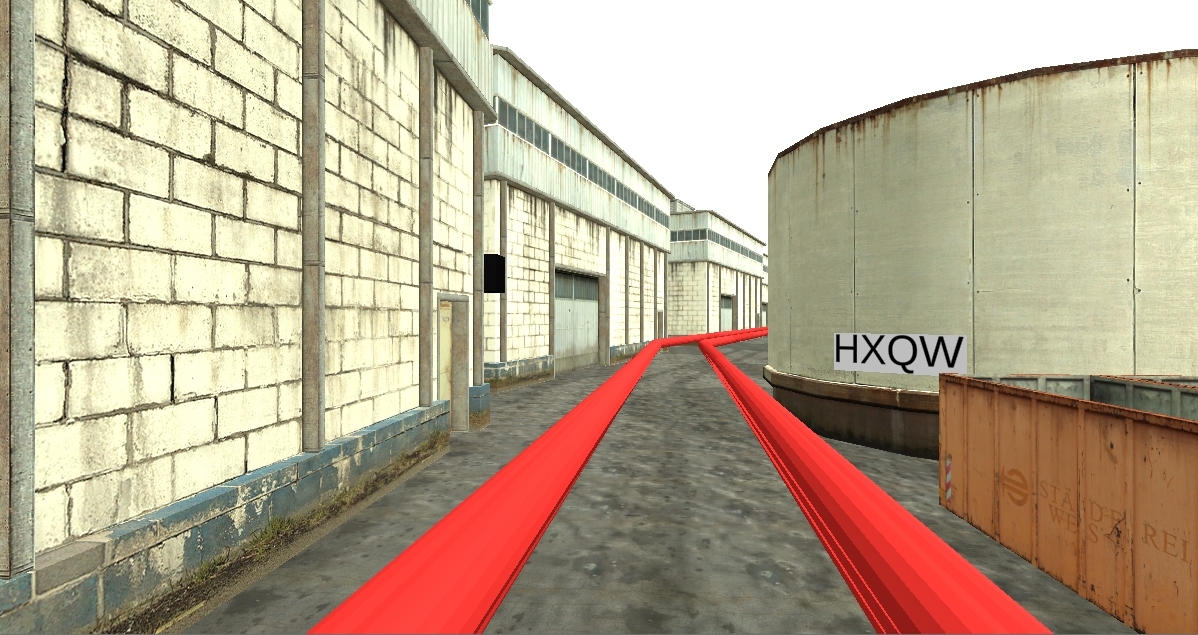}
\caption{} \label{fig:noeffectrec}
\end{subfigure}
\begin{subfigure}{0.48\textwidth}
\includegraphics[width=\textwidth]{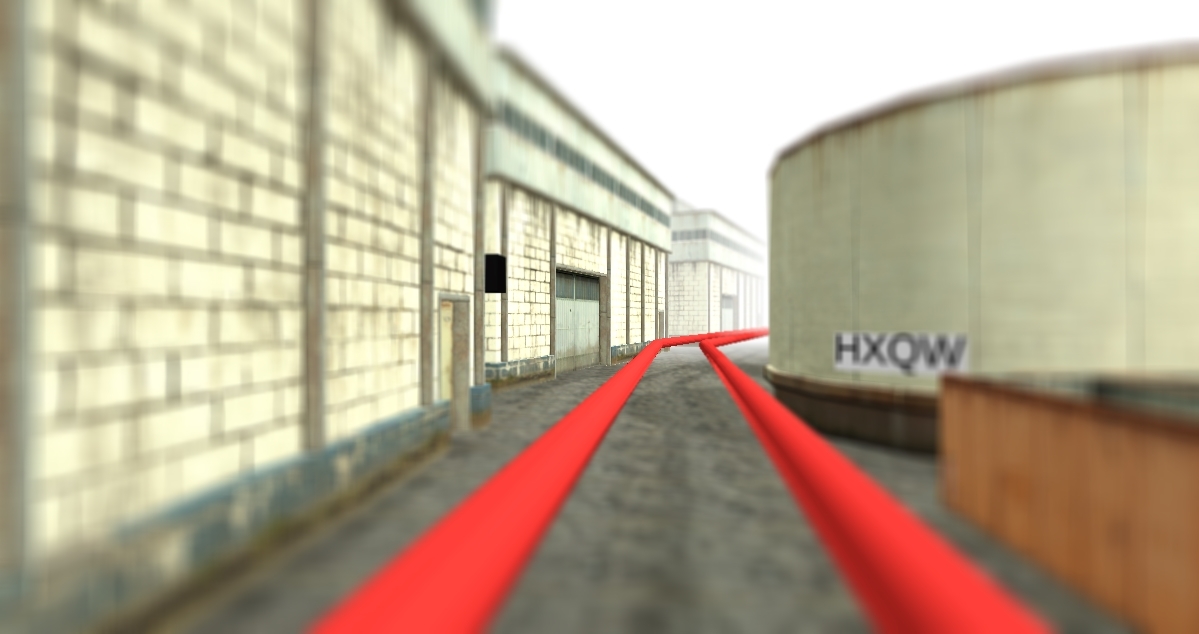}
\caption{} \label{fig:simblurrec}
\end{subfigure}
\caption{The three rendering methods used in the study. (a) (Real) foveated: the user is looking at the visual target on the right (in focus), and the rest of the image is blurred. (b) No effects: everything is in focus. (c) Marginal blur (simulated foveated): the gaze is assumed in the center of the image, and pixels far from it, including the visual target, are blurred }\label{fig:threeperspectives}.
\end{figure} 
In this work, we explicitly investigated the differences between real and simulated foveated blurring rendering while moving on a roller coaster in an immersive VR environment. Figure \ref{fig:threeperspectives} exemplifies the effect seen by users in the two foveated conditions with respect to a view when no effects have been applied. To engage participants in being actively scanning the environment, we asked them to perform a recognition task. We did not expect improvements in the recognition task since it was just designed to engage participants visually. The blurring effect was not so strong to prevent recognition of the surroundings. 

We expected that both real and simulated foveated blurring reduced in some respect visual discomfort or the task difficulty and real foveated being better than the simulated one. Specifically, we measured visual discomfort using a reduced version of the validated questionnaire proposed by Sevinc and colleagues \cite{sevinc2020psychometric} that consider three scales for nausea, oculomotor fatique and disorientation. 


\section{Related work}
A large amount of research work has been dedicated to the study of visually-induced motion sickness (VIMS) \cite{kennedy2010research,saredakis2020factors}. Several authors agree on the necessity of finding good strategies to reduce this effect in virtual environments, and different approaches based on specifically designed rendering solutions have been proposed.

Carnegie and Rhee \cite{carnegie2015reducing} proposed a dynamic depth of field blurring to reduce the amount of perceived motion showing a reduction of nausea and disorientation discomfort in a user study.

Fernandes and Feiner \cite{fernandes2016combating} proposed a dynamic reduction of the field of view that seems to help the participants of a user study to stay in the Virtual Environment longer and feel more comfortable than they did in the control condition.

Buhler et al. \cite{buhler2018reducing} proposed two visual effects: the  “circle effect” (blending the view of a different static camera outside a circle) and the  “dot effect”, (adding rendered dots in peripheral vision that counteracts the virtual motion of the environment when the user is moving), to alter the visualization in the peripheral area of the perceived visual image aiming at reducing effective optical flow. The study did not show significant effects for reducing VR sickness due to the large variability of the effects.

Cao et al. \cite{cao2018visually} proposed the addition of static rest frames and dynamic rest frames (with velocity-dependent opacity) to limit sickness.
Nie et al. \cite{nie2019analysis} used adaptive blurring of the background. The method detects not salient regions and applies selective blurring only to them.

FOV reduction, DOF blurring, and rest-frame have been compared in a racing game task in \cite{shi2021virtual}.
A comparison of some of those techniques is proposed in \cite{choros2019software}. The study suggested a weaker effect of the rest frame with respect to other methods.

A selection of the methods has also been implemented in a Unity toolkit and made available in an open-source repository \cite{ang2020gingervr}.

As suggested in \cite{pai2016gazesim}, gaze estimation might be an effective approach to select the regions where applying optical flow reduction.
The use of eye-tracking to control sickness reduction algorithms has been proposed first in \cite{adhanom2020effect}. In this work, the authors compare foveated vs non-foveated large FOV reductions without specific tasks performed by the users.

In this work, we aimed to assess the effect on real vs. simulated foveated blurring empirically. Our study complement and extend the results provided by  Carnegie and Rhee \cite{carnegie2015reducing} on foveated blurring while offering empirical support to the approach of using "real" foveated as initially proposed by Pai and colleagues \cite{pai2016gazesim}.

\section{VR environment and task}
\begin{figure}
\centering
\includegraphics[width=0.6\textwidth]{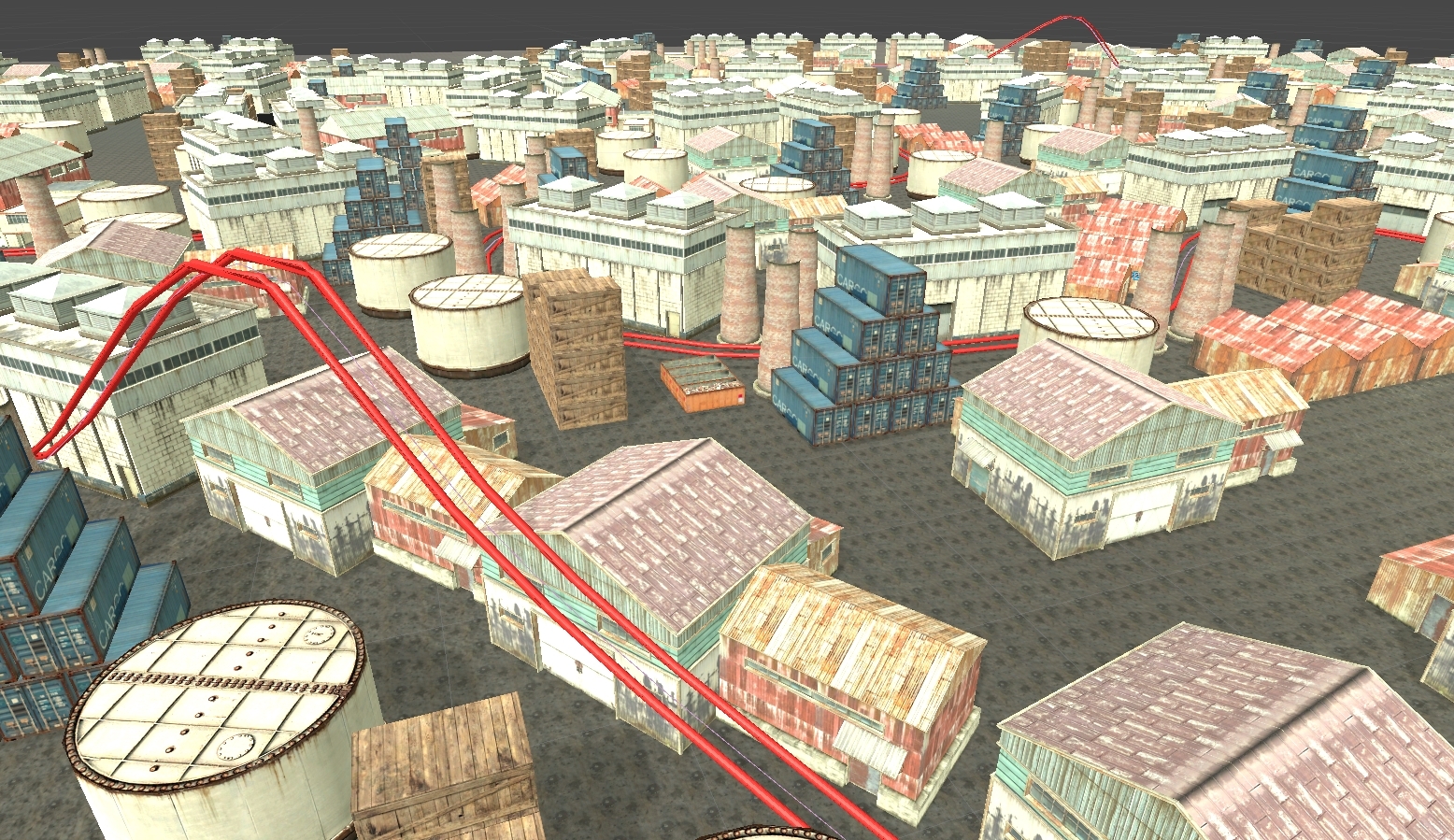}
\caption{View of the roller coaster environment. The user position is anchored to a cart moving on the red rail track surrounded by heavily textured buildings.}
\label{fig:trackview}
\end{figure}

The VR environment used for the study realizes a roller coaster simulation with a heavily textured scene with a relevant optical flow at the margins of the visual field to foster sickness (Figure \ref{fig:trackview}). This setting has been chosen because of its natural form of constrained navigation, limiting possible biases related to user control.

The simulation program was implemented in the Unity framework. The VR experience was created to use HTC Vive Pro Eye Head-Mounted Display (HMD) equipped with a 120Hz eye-tracking technology. In each ride, the user goes through a VR roller coaster experience in which they are sitting in a cart, moving automatically through the course that consisted of a set of rails organized in a closed loop. The course features a variety of turns, uphill and downhill sections, and speed variations introduced to induce motion sickness. A full lap around the course track takes about 160 seconds.

The simulation can be executed with three different visualization modes:
 \textit{foveated blurring}, in which only a circular area is in focus (centered on the actual user's gaze point as tracked by an eye-tracking device embedded in the HMD) while the rest of the image is blurred; \textit{simulated foveated blurring}, similar to the above but the gaze point is assumed fixed in the middle of the rendered frame; and \textit{no effects}, on which no blurring effects are applied and all the scene in focus.
 
For both the blurring modes, the in-focus area is a circle with radius \textit{r} while, outside of it, a Gaussian blur with standard deviation $\sigma$ is applied. The effect is smoothed from the edge of the in-focus circle with a ramp up parameter that scales linearly with the distance from the center of the circle. The parameters \textit{r} and $\sigma$ have been tuned in a set of pilot studies with participants not involved in the main study.

A set of 35 visual targets (similar to those visible in Figure \ref{fig:threeperspectives}) are placed at different heights and angles on the sides of the track. Each one displays a string of 4 letters. Five of the targets contain the "X" character, and they have to be identified by the users while riding the roller coaster. This recognition task is relatively easy, but it requires active visual search and attention.

Acknowledging the limitations of the eye-tracking device\cite{sipatchin2020accuracy}, the recognition task has been designed in such a way that visual search does not require frequent and fast head movements.

\section{The study}
\label{sec:study}
The study design was a full factorial experiment with three within-subject conditions based on the visualization modes described above: (i) \textit{(Real) Foveated}, 
 (ii) \textit{Simulated foveated}, and 
(iii) \textit{No effects} 
The study was carried out in the VIPS lab at the University of Verona  and received ethical clearance by the Human Research Approval Committee of the same university.
The hypotheses were the following:
\begin{itemize}
\item \textit{H1}: in general, foveated blurring reduces visual discomfort and the (perceived) task difficulty, with respect to the \textit{no effects} condition (as partially demonstrated by Carnegie \& Rhee \cite{carnegie2015reducing});
\item \textit{H2}: the real foveated blurring technique reduces visual discomfort and task difficulty with respect to simulated blurring (as proposed, without empirical evidence, by Pai et al. \cite{pai2016gazesim})
\end{itemize}


\subsection{Participants and procedure}
Thirty-eight (38) participants have been recruited (20 females with age between 18 to 25, 18 males with age between 18 to 25, and 2 males with age between 26 to 30). All of them had normal or corrected to normal vision. Twenty (20) participants had no previous experience with VR, and  13 had only one previous experience. 

Each participant was tested individually. Upon arrival, s/he received the informed consent form, the privacy form, and the information about the COVID-19 procedure adopted. S/he was also informed that the experiment can be interrupted at any moment (nobody actually did) and was asked to fill the pre-task questionnaire. 

Then, the participant was invited to wear the HMD and perform the eye tracker's calibration. This procedure provided an initial exposition to VR for the participants without previous experience and assessed their willingness to continue the study.

The participant was then assigned to a randomized order and repeated the roller coaster experience in the three conditions. The positions of the targets to recognize were also randomized within the conditions and the participants. 
Each condition involved one single ride in the roller coaster in one of the visualization modes described above, followed by a 3-minutes break before starting the next ride. For the recognition task, the participants had to verbally report to the experimenter when they located a target with an "X" inside. 

The participant was eventually thanked and dismissed. No compensation was provided for the participation.

\subsection{Measurements}

A pre-task questionnaire was used to collect demographic information (gender and age), previous experience with VR, self-reported vision problems, self-reported sensitivity to sickness (measured on a 4-item scale from never to very often).

Measures of visual discomfort and task difficulty have been collected after each condition.
For measuring visual discomfort, a reduced version of the Simulator Sickness Questionnaire (SSQ) \cite{sevinc2020psychometric} has been used. Specifically, we used the items General discomfort, Fatigue, Eyestrain, Difficulty focusing, Sweating, Nausea, Blurred vision (all 4-point Likert scales from none to severe) which correspond to the scales of Nausea, Oculomotor, and Disorientation. 

In order to measure the perceived task difficulty,  a single item on a 4-point Likert scale (from very simple to very difficult) has been used.

\section{Results}
\label{sec:results}
Although the variability in the data is very high (see Table \ref{tab:res}), the three SSQ scales have a good level of reliability (their Cronbach’s alpha values are $0.71$ for the \textit{Nausea} scale, $0.84$ for the \textit{Oculomotor} scale and $0.76$ for the \textit{Disorientation scale}). It is worth noting that at least part of the variability is due to the self-declared sensitivity to sickness: its Pearson correlation with the \textit{Nausea} scale is $r=0.48$, $p<0.01$; with the \textit{Oculomotor} scale is $r=0.56$, $p<0.01$ and with the \textit{Disorientation} scale is $r=0.59$, $p<0.01$. 

A first attempt to fit a Linear Mixed Effect Model on each scale by using both condition and the pre-test variables of gender, sight, and sensitivity to sickness as fixed effects with the individual participants as random effects confirms a statistically significant effect of sickness on each scale and also an effect of condition on the \textit{Nausea} and \textit{Disorientation} scales (but not for \textit{Oculomotor scale}). Neither gender nor sight have statistically significant effects.  

Friedman non-parametric tests (using the individuals as grouping variable) revealed a difference between the three conditions on the \textit{Nausea} scale (Friedman, $f=10.84$, $p<0.01$), and  Conover post-hoc tests confirm a significant difference between the \textit{no effects condition} and the other two (respectively $p=0.0054$ for \textit{foveated} vs \textit{no effects} and $p=0.0054$ for \textit{simulated} vs \textit{no effects}) but no difference between \textit{foveated} and \textit{simulated}. No statistical differences were found for the other two scales (respectively, Friedman $f=1.54$, $p=0.46$ for \textit{Oculomotor} and Friedman $f=3.36$, $p=0.19$ for \textit{Disorientation}).  A significant effect has been found for the difficulty of task (Friedman $f=12.12$, $p<0.01$). Conover post-hoc tests confirm a difference between \textit{foveated} and \textit{simulated} conditions ($p<0.01$) and between \textit{no effects} and \textit{simulated} ($p<0.01$) but not between \textit{foveated} and \textit{no effects}. 

To better discount the effect of self-declared sensitivity to sickness, we decided to categorize sickness as low when the value of self-declared sensitivity to sickness was equal to 1 and high for the values 2,3 and 4. This procedure partitioned the data into two groups of 21 participants with \textit{low sensitivity to sickness} and 17 participants with \textit{high sensitivity to sickness}.

Among the participants with \textit{low sensitivity to sickness}, a Friedman non-parametric tests (using the individuals as grouping variable) revealed a difference among the three conditions for the \textit{Nausea} scale (Friedman, $f=9.29$, $p<0.01$), and  Conover post-hoc tests confirm a significant difference between the \textit{no effects} condition and the other two (respectively $p<0.05$ for \textit{foveated} vs \textit{no effect} and $p<0.01$ for \textit{simulated} vs \textit{no effects}) but no difference between \textit{foveated} and \textit{simulated}. No statistical differences were found for the other two scales (respectively, Friedman $f=0.12$, $p=0.94$ for \textit{Oculomotor} and Friedman $f=0.047$, $p=0.98$ for \textit{Disorientation}).  A significant effect has been found for \textit{difficulty of task} (Friedman $f=8.16$, $p=0.01$). Conover post-hoc tests confirm a difference between \textit{foveated} and \textit{simulated} conditions and between \textit{no effects} and \textit{simulated} but not between \textit{foveated} and \textit{no effects}. 

For the participants with (self-declared) \textit{high sensitivity on sickness}, there is no difference for the \textit{Nausea} scale (Friedman $f=2.63$, $p=0.27$) nor for the \textit{Oculomotor} scale (Friedman $f=4.38$, $p=0.11$) and the \textit{task difficulty} item (Friedman $f=4.8$, $p=0.09$). There is a statistically significant effect for the \textit{Disorientation} scale (Friedman $f=7.66$, $p<0.05$). Conover post-hoc tests reveal a difference between \textit{foveated} and \textit{simulated} but not between those two and no effect.
As expected, there are no differences in the recognition task: $94\%$ of the participants recognized more than 4 targets and only 5 participants recognized 3 or less.

\begin{table}[t]
\centering
\begin{tabular}{l|rrr|l|rrr|l|rrr|l|rrr|}
\cline{2-4} \cline{6-8} \cline{10-12} \cline{14-16}
 & \multicolumn{3}{c|}{Scale Nausea} &  & \multicolumn{3}{c|}{Scale Oculomotor} &  & \multicolumn{3}{c|}{Scale Disorient.} &  & \multicolumn{3}{c|}{Task difficulty} \\ \cline{2-4} \cline{6-8} \cline{10-12} \cline{14-16} 
 & \multicolumn{1}{c|}{all} & \multicolumn{1}{c|}{\begin{tabular}[c]{@{}c@{}}low\\ sick\end{tabular}} & \multicolumn{1}{c|}{\begin{tabular}[c]{@{}c@{}}high\\ sick\end{tabular}} &  & \multicolumn{1}{c|}{all} & \multicolumn{1}{c|}{\begin{tabular}[c]{@{}c@{}}low\\ sick\end{tabular}} & \multicolumn{1}{c|}{\begin{tabular}[c]{@{}c@{}}high\\ sick\end{tabular}} &  & \multicolumn{1}{c|}{all} & \multicolumn{1}{c|}{\begin{tabular}[c]{@{}c@{}}low\\ sick\end{tabular}} & \multicolumn{1}{c|}{\begin{tabular}[c]{@{}c@{}}high\\ sick\end{tabular}} &  & \multicolumn{1}{c|}{all} & \multicolumn{1}{c|}{\begin{tabular}[c]{@{}c@{}}low\\ sick\end{tabular}} & \multicolumn{1}{c|}{\begin{tabular}[c]{@{}c@{}}high\\ sick\end{tabular}} \\ \cline{1-4} \cline{6-8} \cline{10-12} \cline{14-16} 
\multicolumn{1}{|l|}{\multirow{2}{*}{foveated}} & 1.90 & 1.74 & 2.92 &  & 2.02 & 1.87 & 3.00 &  & 1.83 & 1.68 & \textbf{2.83} &  & 2.07 & 1.96 & 2.50 \\
\multicolumn{1}{|l|}{} & (.74) & (.67) & (.17) &  & (.67) & (.53) & (.41) &  & (.72) & (.56) & \textbf{(.88)} &  & (.69) & (.66) & (.50) \\ \cline{1-4} \cline{6-8} \cline{10-12} \cline{14-16} 
\multicolumn{1}{|l|}{\multirow{2}{*}{no effects}} & \textbf{2.07} & \textbf{1.91} & 3.08 &  & 2.01 & 1.85 & 3.06 &  & 1.89 & 1.72 & 3.00 &  & 2.13 & 2.03 & 2.25 \\
\multicolumn{1}{|l|}{} & \textbf{(.79)} & \textbf{(.70)} & (.57) &  & (.71) & (.55) & (.83) &  & (.71) & (.54) & (.72) &  & (.63) & (.60) & (.50) \\ \cline{1-4} \cline{6-8} \cline{10-12} \cline{14-16} 
\multicolumn{1}{|l|}{\multirow{2}{*}{simulated}} & 1.92 & 1.74 & 3.08 &  & 2.08 & 1.88 & 3.38 &  & 1.96 & 1.78 & \textbf{3.08} &  & \textbf{2.40} & \textbf{2.31} & 3.00 \\
\multicolumn{1}{|l|}{} & (.77) & (.66) & (.32) &  & (.77) & (.58) & (.60) &  & (.74) & (.60) & \textbf{(.63)} &  & \textbf{(.77)} & \textbf{(.74)} & (.82) \\ \cline{1-4} \cline{6-8} \cline{10-12} \cline{14-16} 
\end{tabular}
\vspace{1em}
\caption{Means and standard deviations in the different conditions and levels of sensitivity of sickness. The values displayed in bold have statistical significant difference with the others across the same condition.} \label{tab:res}
\end{table}


\section{Discussion}
\label{sec:disc}

Our results, although still preliminary, provide evidence that both techniques of foveated blurring reduce visual discomfort (hypothesis H1), specifically for nausea, while the real foveated reduces the perceived task difficulty (hypothesis H2), but only for users with low sensitivity to sickness. Participants with high sensitivity for sickness do not seem to benefit from the blurring techniques for reducing nausea or the perceived task difficulty. Still, we found a small effect on reducing disorientation.

Although still preliminary, our results consider the moderating effect of individual sensitivity to sickness and provide some evidence that it plays a role in the foveated rendering effect. In this respect, it contributes with new evidence to the growing body of research. Indeed, the variations in the scores, although small, were, at least for the \textit{Nausea} and \textit{Disorientation} scales in line with those found in other studies \cite{adhanom2020effect},\cite{carnegie2015reducing}. 

Differently from previous works, we had noticeable effect only on the \textit{Nausea} and \textit{Disorientation} scale and not on the \textit{Oculomotor} scale for the SSQ.  That might be due to the constrained navigation or the short duration of the ride that could have limited the VR experience's sickness effect. Another explanation might be that our task required more cognitive attention than simple navigational tasks used in other studies, and that may have an impact on the perception of ocular strain\cite{iskander18}.

Another novel result is that we also measured the effect on the perceived task difficulty and observed a small but significant effect, particularly in differentiating real from simulated foveated blurring. 

Our experiment had some weaknesses and limitations that we plan to overcome in future studies. The single session duration was perhaps too short to create a relevant sickness on many subjects. The number of subjects tested was higher than in other studies but still limited;. It would be better to have a larger number of participants would allow to usinge more powerful statistical tools. Both the sensitivity level of sickness and the experienced sickness were self-reported while in further studies they should be objectively measured. 

For active recognition, it would also be helpful to use a more challenging task. Although we still believe that an active task helps control the participants' engagement with the environment, our task was probably too easy. 

Finally, an important issue to be considered is related to the accuracy and the latency of the eye tracker. Recent studies \cite{sipatchin2020accuracy} revealed that the device we used is less precise and accurate far from the center of the headset FOV. The latency of the tracking could, in principle, also limit the sickness reduction.

\section{Conclusion}
\label{sec:disc}
In conclusion, our results suggest that for participants with higher self-declared sensitivity to sickness, we found some evidence that the foveated technique provides a small benefit on disorientation for simulated blurring but not enough to improve with respect to applying no effects. For participants with lower sensitivity to sickness, there might be an improvement for nausea when using blurring. Still, the specific technique does not matter, although the (perceived) difficulty of the task seems to improve when the real foveated method is used. 

The effects of the differences are minimal (and the sample is small); therefore, the present study results should be considered an indication for further research rather than robust evidence. Nevertheless, we believe that this study contributes to the ongoing discussion on the effects of foveated techniques and, in particular, on the opportunity to implement "real" foveated using eye-tracking. The difference found in our study between participants with different levels of sensitivity to sickness has never been reported in the literature and deserves further investigation.

%
%
\bibliographystyle{splncs04}
\bibliography{samplepaper}

\begin{thebibliography}{10}
\providecommand{\url}[1]{\texttt{#1}}
\providecommand{\urlprefix}{URL }
\providecommand{\doi}[1]{https://doi.org/#1}

\bibitem{adhanom2020effect}
Adhanom, I.B., Griffin, N.N., MacNeilage, P., Folmer, E.: The effect of a
  foveated field-of-view restrictor on vr sickness. In: 2020 IEEE conference on
  virtual reality and 3D user interfaces (VR). pp. 645--652. IEEE (2020)

\bibitem{ang2020gingervr}
Ang, S., Quarles, J.: Gingervr: An open source repository of cybersickness
  reduction techniques for unity. In: 2020 IEEE Conference on Virtual Reality
  and 3D User Interfaces Abstracts and Workshops (VRW). pp. 460--463. IEEE
  (2020)

\bibitem{buhler2018reducing}
Buhler, H., Misztal, S., Schild, J.: Reducing vr sickness through peripheral
  visual effects. In: 2018 IEEE Conference on Virtual Reality and 3D User
  Interfaces (VR). pp. 517--9. IEEE (2018)

\bibitem{cao2018visually}
Cao, Z., Jerald, J., Kopper, R.: Visually-induced motion sickness reduction via
  static and dynamic rest frames. In: 2018 IEEE conference on virtual reality
  and 3D user interfaces (VR). pp. 105--112. IEEE (2018)

\bibitem{carnegie2015reducing}
Carnegie, K., Rhee, T.: Reducing visual discomfort with hmds using dynamic
  depth of field. IEEE computer graphics and applications  \textbf{35}(5),
  34--41 (2015)

\bibitem{choros2019software}
Choro{\'s}, K., Nippe, P.: Software techniques to reduce cybersickness among
  users of immersive virtual reality environments. In: Asian Conference on
  Intelligent Information and Database Systems. pp. 638--648. Springer (2019)

\bibitem{fernandes2016combating}
Fernandes, A.S., Feiner, S.K.: Combating vr sickness through subtle dynamic
  field-of-view modification. In: 2016 IEEE symposium on 3D user interfaces
  (3DUI). pp. 201--210. IEEE (2016)

\bibitem{iskander18}
Iskander, J., Hossny, M., Nahavandi, S.: A review on ocular biomechanic models
  for assessing visual fatigue in virtual reality. IEEE Access  \textbf{6},
  19345--19361 (2018)

\bibitem{kennedy2010research}
Kennedy, R.S., Drexler, J., Kennedy, R.C.: Research in visually induced motion
  sickness. Applied ergonomics  \textbf{41}(4),  494--503 (2010)

\bibitem{nie2019analysis}
Nie, G.Y., Duh, H.B.L., Liu, Y., Wang, Y.: Analysis on mitigation of visually
  induced motion sickness by applying dynamical blurring on a user's retina.
  IEEE transactions on visualization and computer graphics  \textbf{26}(8),
  2535--2545 (2019)

\bibitem{pai2016gazesim}
Pai, Y.S., Tag, B., Outram, B., Vontin, N., Sugiura, K., Kunze, K.: Gazesim:
  simulating foveated rendering using depth in eye gaze for vr. In: ACM
  SIGGRAPH 2016 Posters, pp.~1--2 (2016)

\bibitem{patney2016towards}
Patney, A., Salvi, M., Kim, J., Kaplanyan, A., Wyman, C., Benty, N., Luebke,
  D., Lefohn, A.: Towards foveated rendering for gaze-tracked virtual reality.
  ACM Transactions on Graphics (TOG)  \textbf{35}(6),  1--12 (2016)

\bibitem{sagnier2020user}
Sagnier, C., Loup-Escande, E., Lourdeaux, D., Thouvenin, I., Vall{\'e}ry, G.:
  User acceptance of virtual reality: an extended technology acceptance model.
  International Journal of Human--Computer Interaction  \textbf{36}(11),
  993--1007 (2020)

\bibitem{saredakis2020factors}
Saredakis, D., Szpak, A., Birckhead, B., Keage, H.A., Rizzo, A., Loetscher, T.:
  Factors associated with virtual reality sickness in head-mounted displays: a
  systematic review and meta-analysis. Frontiers in human neuroscience
  \textbf{14} (2020)

\bibitem{sevinc2020psychometric}
Sevinc, V., Berkman, M.I.: Psychometric evaluation of simulator sickness
  questionnaire and its variants as a measure of cybersickness in consumer
  virtual environments. Applied ergonomics  \textbf{82},  102958 (2020)

\bibitem{shi2021virtual}
Shi, R., Liang, H.N., Wu, Y., Yu, D., Xu, W.: Virtual reality sickness
  mitigation methods: A comparative study in a racing game. arXiv preprint
  arXiv:2103.05200  (2021)

\bibitem{sipatchin2020accuracy}
Sipatchin, A., Wahl, S., Rifai, K.: Accuracy and precision of the htc vive pro
  eye tracking in head-restrained and head-free conditions. Investigative
  Ophthalmology \& Visual Science  \textbf{61}(7),  5071--5071 (2020)

\end{thebibliography}

\end{document}